\documentclass[reprint,aps,pra,twocolumn,showpacs,amsmath,amssymb,floatfix]{revtex4-1}

\usepackage{mathptmx}
\usepackage{float}
\usepackage{epsfig}
\usepackage{amsmath}
\usepackage{amssymb}\vskip4mm
\usepackage{bm}
\usepackage{graphicx}
\usepackage{color}
\usepackage{rotating}
\usepackage{color}

\begin{document}
\title{Bessel-Bessel laser bullets doing the twist}

\author{Yousef I. Salamin }
\email{ysalamin@aus.edu}

\affiliation{Max-Planck-Institut f\"{u}r Kernphysik, Saupfercheckweg 1, 60117 Heidelberg, Germany,}
\affiliation{Department of Physics, American University of Sharjah, POB 26666, Sharjah, United Arab Emirates}

\date{\today}

\begin{abstract}
Bessel beams carry orbital angular momentum (OAM). Opening up of the Hilbert space of OAM for information coding makes Bessel beams potential candidates for utility in data transfer and optical communication. A laser bullet is the ultra-short and tightly-focused analogue of a non-diffracting and non-dispersing laser Bessel beam. Here, we show fully analytically that a Bessel-Bessel laser bullet possesses orbital angular momentum. Analytic investigation of the energy, linear momentum, energy flux, and angular momentum, associated with the fields of a Bessel-Bessel bullet, in an under-dense plasma, is conducted. The expressions reported here will play a crucial role in preparing the laser bullets for practical applications, such as data transfer in optical communication, x- and gamma-ray generation from colliding bullets with counter-propagating electron bunches, particle trapping, tweezing and laser acceleration.
\end{abstract}

\maketitle

\section{Introduction}

In addition to being diffraction- and dispersion-free, Bessel beams carry orbital angular momentum (OAM)  \cite{padgett,bliokh,schulze}, making them ideal for many applications, such as high-resolution microscopy \cite{kozawa1}, optical trapping and tweezing \cite{yao}, precision drilling \cite{kozawa2,duocastella} and laser acceleration \cite{salamin-pla}. Opening up of the Hilbert space of OAM to information coding makes Bessel beams potential candidates for utility in data transfer and optical communications \cite{dudley}. 

The ultra-short (temporally or spatially) and tightly-focused analogue of a non-diffracting and non-dispersing laser pulse is often referred to as a laser bullet \cite{durnin1,durnin2,zhong,trapani,siviloglou}. Such objects have attracted considerable attention over the past decade or so \cite{chong,naidoo,zong,urrutia,mendoza,volke}. For example, Airy-Bessel bullets \cite{chong} and higher-order Poincar\'e sphere beams \cite{naidoo} have been suggested and realized in laboratory experiments. 

A new addition to the list of light bullets is the so-called Bessel-Bessel bullet \cite{salamin-oe,salamin-sr}. Analytic expressions for the electric and magnetic fields of a Bessel-Bessel bullet, propagating in an under-dense plasma, of plasma frequency $\omega_p$, have recently been derived. The fields stem from the zeroth-order vector potential (SI units and circular-cylindrical coordinates, $r$, $\theta$ and $z$, are used throughout)
\begin{equation}\label{A}
 	A(r,\theta,\eta,\zeta) = a_0J_l(k_rr) j_0\left(\frac{\pi\zeta}{L}\right) e^{i(\varphi_0+k_0\zeta+l\theta-\alpha\eta)},
\end{equation}
where $\eta = (z+ct)/2$, $\zeta = z-ct$, $c$ is the speed of light in vacuum, $a_0$ is a constant amplitude, $J_l$ and $j_0$ are, respectively, ordinary and spherical Bessel functions of their given arguments and integer orders $l$ and zero, respectively. Furthermore, $\varphi_0$ is a constant initial phase and $k_0 = 2\pi/\lambda_0$ is a central wavenumber for the pulse, corresponding to the central wavelength $\lambda_0$. The bullet described by the vector potential (\ref{A}) is assumed to have an axial (spatial) extension $L\sim c\tau$, where $\tau$ is its temporal full-width-at-half-maximum. On the other hand, a waist radius at focus for the pulse, $w_0$, is determined from $w_0=x_{1,l}/k_r$, where $x_{1,l}$ is the first zero of $J_l$. Finally, in (\ref{A})
\begin{equation}
	\alpha = \frac{k_r^2+k_p^2}{2k_0};\quad k_p = \frac{\omega_p}{c};\quad \omega_p = \sqrt{\frac{n_0e^2}{m\varepsilon_0}},
\end{equation}
with $n_0$ the ambient electron density of the plasma, $\varepsilon_0$ the permittivity of the vacuum, and $m$ and $-e$ the mass and charge, respectively, of the electron. Equation (\ref{A}) has been obtained \cite{salamin-oe,salamin-sr,esarey} from solving the wave equations satisfied by the scalar and vector potentials with inhomogeneous terms which, in turn, stem from interaction of the laser pulse with an under-dense plasma, assumed to be linear, with the two potentials linked by the Lorentz gauge. Explicit expressions for the electric and magnetic field components, derived from (\ref{A}) are given by Eqs. (20)-(24) in \cite{salamin-sr}. Replacing $j_0$ in the amplitude of the vector potential (\ref{A}) with an Airy function brings to it formal resemblance with the field amplitude of an Airy-Bessel bullet \cite{siviloglou,chong}.

\section{Propagation characteristics}

Some of the key propagation characteristics of a Bessel-Bessel bullet may be highlighted, starting with the phase of the vector potential. Since the Bessel functions in Eq. (\ref{A}) are real, the phase of $A$ is fully accounted for by $\varphi = \varphi_0+k_0\zeta+l\theta-\alpha\eta$. For example, the wavevector of the pulse, in circular cylindrical coordinates, may be obtained from \cite{mcdonald1}
\begin{equation}\label{k}
	\bm{k} = \bm{\nabla}\varphi = \left(\frac{l}{r}\right) \hat{\bm{\theta}} +\left(k_0-\frac{\alpha}{2}\right)\hat{\bm{z}},
\end{equation}
where $\hat{\bm{\theta}}$ and $\hat{\bm{z}}$ are unit vectors in the azimuthal and axial directions, respectively. The message of Eq. (\ref{k}) is simple: the normal to a surface of constant phase is not parallel to the direction of propagation, rather it makes an angle, $\beta$, with $\hat{\bm{z}}$ given by
$\tan\beta = (l/r)/(k_0-\alpha/2)$.
This is equivalent to saying that, apart from the $l = 0$ case, a wavefront is not planar and normal to the direction of propagation (like in the case of a plane wave) but is a helix of fixed radius $r$ and axis along $\hat{\bm{z}}$ \cite{mcdonald1,berry}.

\begin{figure}
\centering
\includegraphics[width=6.8cm]{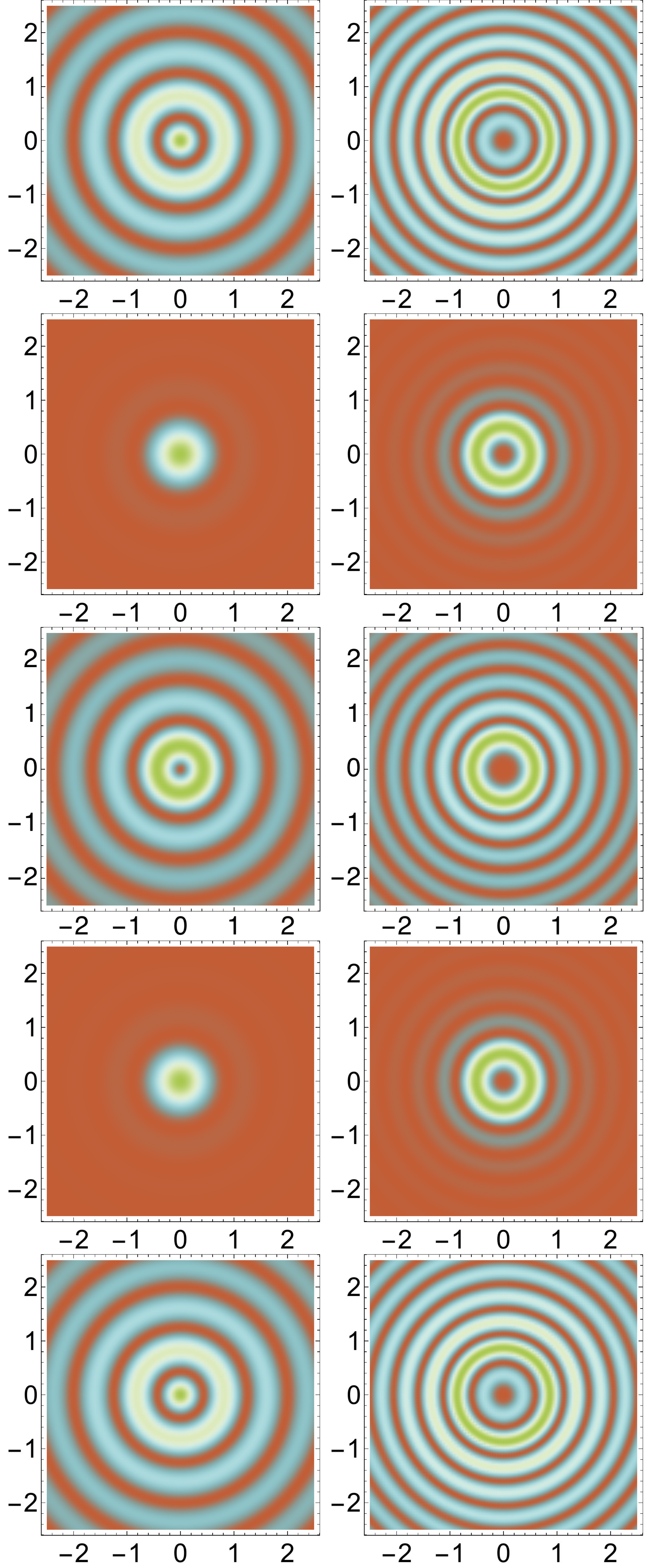}
\begin{picture}(0,0)(0,0)
\put(-148,-10){$x/\lambda_0$}
\put(-52,-10){$x/\lambda_0$}
\put(-205,44){\begin{sideways}$y/\lambda_0$\end{sideways}}
\put(-205,138){\begin{sideways}$y/\lambda_0$\end{sideways}}
\put(-205,232){\begin{sideways}$y/\lambda_0$\end{sideways}}
\put(-205,325){\begin{sideways}$y/\lambda_0$\end{sideways}}
\put(-205,420){\begin{sideways}$y/\lambda_0$\end{sideways}}
\end{picture}
\vskip4mm
\caption{Intensity profiles (top to bottom) of $|E_r/E_0|^2$, $|E_\theta/E_0|^2$, $|E_z/E_0|^2$, $|cB_r/E_0|^2$, and $|cB_\theta/E_0|^2$ in the moving focal plane ($ z = ct, t = 1$ fs) of a Bessel-Bessel bullet for which $L = 1.6 \lambda_0$, $w_0 = 0.9 \lambda_0$, $\lambda_0 = 1 ~\mu$m, in a plasma of electron density $n_0 = 10^{20}$ cm$^{-3}$. Left column: $l = 1$, and right column: $l = 3$. Other parameters used are: $\varphi_0 = 0$, and $k_r = x_{1,l}/w_0$, where $x_{1,l}$ is the first zero of $J_l$. See Figs. \ref{fig3} and \ref{fig4} below for the relative intensities of the various rings displayed here.}
\label{fig1}
\end{figure}
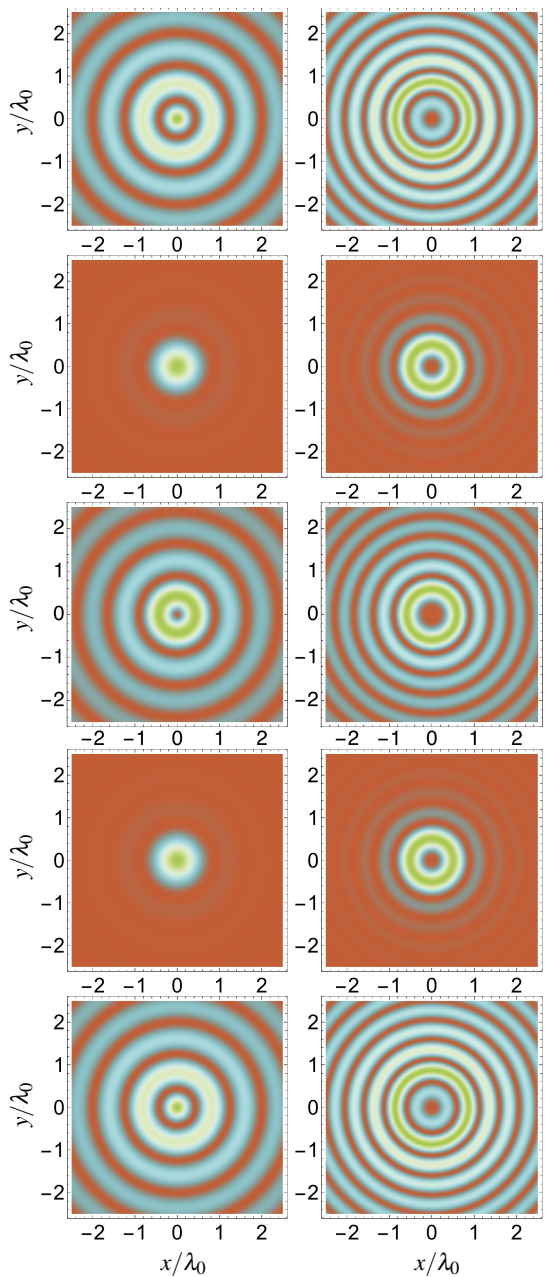

The phase may also be used to derive a dispersion relation. First, an effective frequency is obtained from 
\begin{equation}
	\omega = -\frac{\partial\varphi}{\partial t} = c\left(k_0+\frac{\alpha}{2}\right).
\end{equation}
Likewise, an effective axial wavenumber follows from 
\begin{equation}
	k_z = \frac{\partial\varphi}{\partial z} = k_0-\frac{\alpha}{2},
\end{equation}
which agrees with Eq. (\ref{k}). Employing these results, one gets $(\omega/c)^2-k_z^2 = 2k_0\alpha$, and subsequently the dispersion relation 
\begin{equation}
	\omega = c\sqrt{k_z^2+k_r^2+k_p^2}.
\end{equation}
The group velocity will have an axial component  \cite{esarey}
\begin{equation}
	v_g = \frac{\partial\omega}{\partial k_z} = c\left[\frac{k_0-\alpha/2}{k_0+\alpha/2}\right],
\end{equation}
while the corresponding phase velocity will be given by
\begin{equation}
	v_{ph} = \frac{\omega}{k_z} = c\left[\frac{k_0+\alpha/2}{k_0-\alpha/2}\right].
\end{equation}

Note, at this point, that $v_{ph}v_g = c^2$, and that $v_g < c$, whereas $v_{ph} > c$. These results should strengthen the case for the OAM-carrying pulses as potential candidates for application in data transfer and digital communications \cite{dudley,naidoo,milione}.

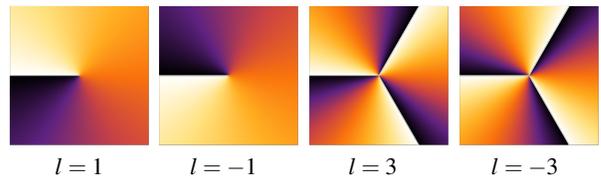
\begin{figure}[t]
\centering
\includegraphics[width=8cm]{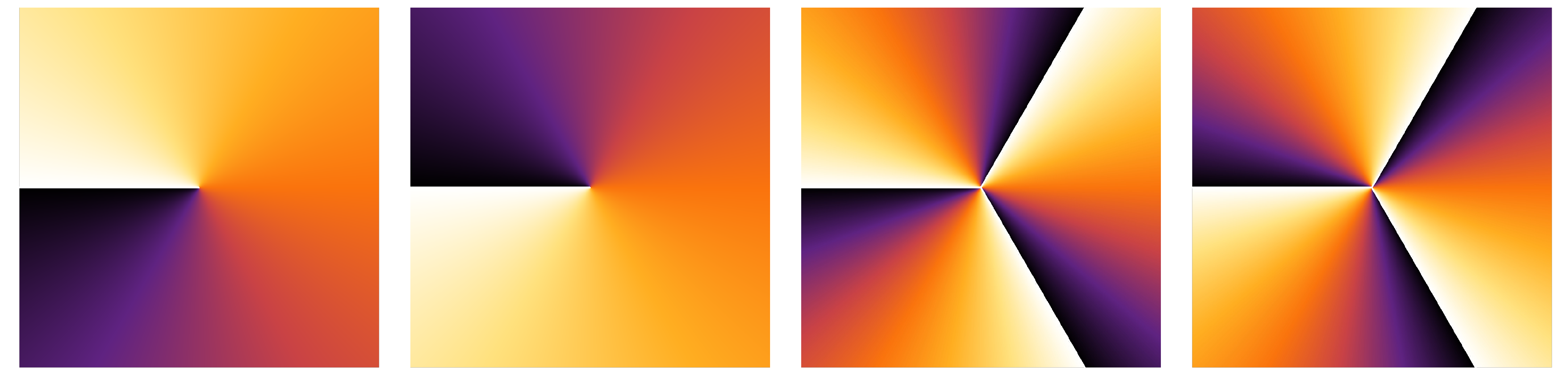}
\begin{picture}(0,0)(0,0)
\put(-210,-10){$l = 1$}
\put(-159,-10){$l = -1$}
\put(-99,-10){$l = 3$}
\put(-45,-10){$l = -3$}
\end{picture}
\vskip4mm
\caption{Density plots of the argument of $e^{i\tilde{\varphi}}$, for four values of the index $l$.}
\label{fig2}
\end{figure}

\section{Fields in the moving focal plane}

Time-dependence of the fields given by Eqs. (20)-(24) in \cite{salamin-sr} is not purely of the form $\exp(\pm i\tilde{\omega} t)$, where $\tilde{\omega}$ is some single frequency. Thus, calculation using those fields of time-averaged quantities ought to start from basic principles, i.e., using the real parts of the fields. This can always be done numerically. Analytically, however, a different approach is needed. 

Recall that our equations represent a pulse that propagates with its shape intact, i.e., diffraction-free and dispersion-free. Recall also the interpretation of $\zeta$ as the coordinate of a point within the pulse relative to its centroid (point of maximum intensity), itself regarded as a {\it moving focal point}. With that in mind, all points on the transverse plane through that centroid always have $\zeta\sim0$, which makes the said transverse plane a {\it moving focal plane} \cite{salamin-oe,salamin-sr,esarey}. In the case of a beam with a stationary focus, the power, for example, is usually calculated by integrating the component of the Poynting vector in the direction of propagation over the entire focal plane, and subsequently averaging the result over time. By analogy, the power carried by our short pulse can be calculated using the fields in the moving focal plane, to be obtained from Eqs. (20)-(24) of \cite{salamin-sr} in the limit of $\zeta\to0$. This procedure, if acceptable for calculation of the time-averaged Poynting vector, may be applied equally as well for calculating other time-averaged quantities, as will be done shortly below. 

From this point onward, arguments of all Bessel functions ($k_rr$) will be suppressed, for convenience. By taking the appropriate limits of Eqs. (20)-(24) of Ref. \cite{salamin-sr}, the electric field components on the moving focal plane of the Bessel-Bessel bullet become
\begin{equation}\label{Er}	
	E_r = E_0\left(\frac{k_r }{k_0}\right) \left(\frac{\alpha-2k_0}{\alpha+2k_0} \right) \left[\frac{J_{l-1} - J_{l+1}}{2}\right] e^{i\tilde{\varphi}},
\end{equation}
\begin{equation}\label{Etheta} 
	E_{\theta} = E_0l \left(\frac{k_r }{k_0}\right) \left(\frac{\alpha-2k_0}{\alpha+2k_0}\right) \left[\frac{J_l}{k_rr}\right] e^{i(\tilde{\varphi}+\pi/2)},
\end{equation}
\begin{equation}
	\label{Ez} E_z = E_0\left(\frac{4\alpha}{\alpha+2k_0} \right)\left[1-\frac{(\pi/L)^2}{3k_0(\alpha+2k_0)}\right] J_le^{i(\tilde{\varphi}+\pi/2)},
\end{equation}
where $\tilde{\varphi} = \varphi_0+l\theta-\alpha\eta$. The associated magnetic field components, on the other hand, take on the following limiting forms
\begin{equation}
	\label{Br} B_r = \frac{E_0}{c} l \left(\frac{k_r }{k_0}\right) \left[\frac{J_l}{k_rr}\right] e^{i(\tilde{\varphi}+\pi/2)},
\end{equation}
\begin{equation}
	\label{Btheta} B_{\theta} = \frac{E_0}{c}\left(\frac{k_r }{k_0}\right) \left[\frac{J_{l-1} - J_{l+1}}{2}\right] e^{i(\tilde{\varphi}+\pi)}. 
\end{equation}

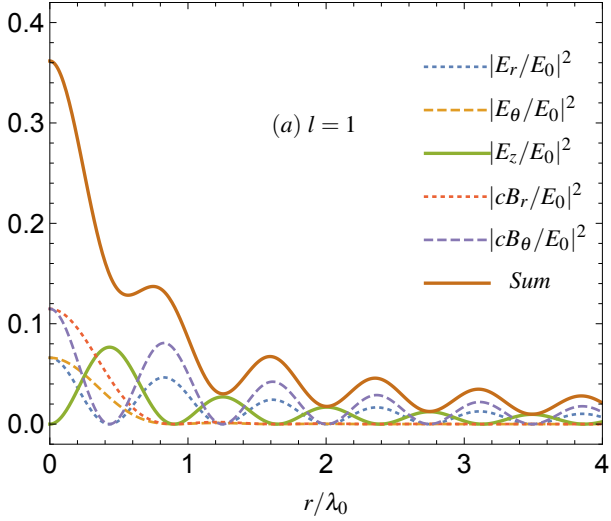
\begin{figure}[t]
\centering
\includegraphics[width=8cm]{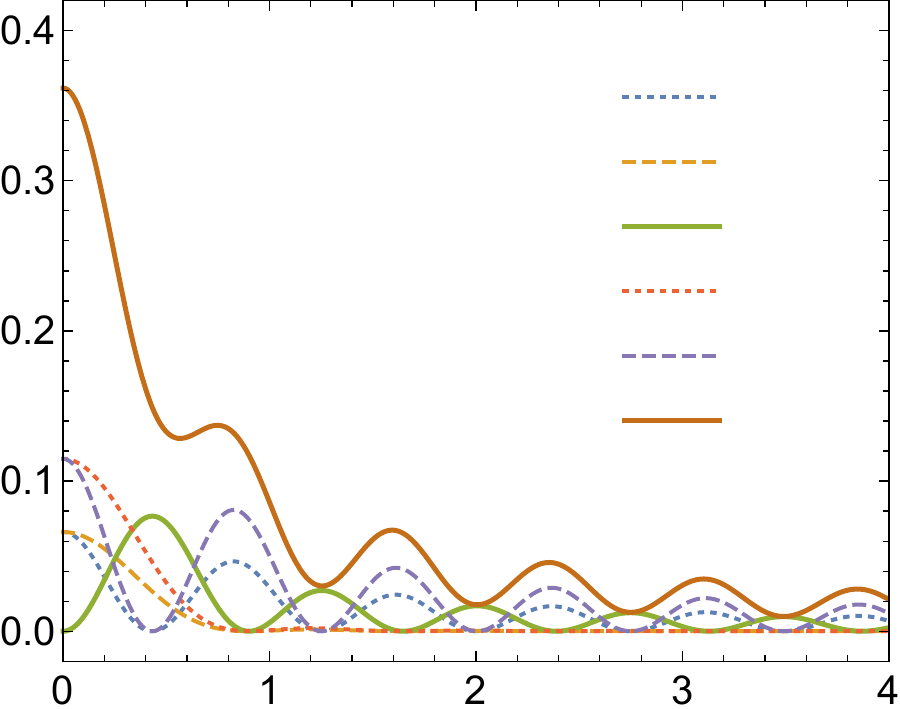}
\begin{picture}(0,0)(0,0)
\put(-118,-13){$r/\lambda_0$}
\put(-48,153){$|E_r/E_0|^2$}
\put(-48,136){$|E_\theta/E_0|^2$}
\put(-48,120){$|E_z/E_0|^2$}
\put(-48,103){$|cB_r/E_0|^2$}
\put(-48,87){$|cB_\theta/E_0|^2$}
\put(-40,71){$Sum$}
\put(-130,130){$(a) ~l = 1$}
\end{picture}
\vskip4mm
\caption{Intensity profiles associated with the field components in the moving focal plane, for the case of $l = 1$, as functions of the radial distance from the focus, using parameters the same as in Fig. \ref{fig1}.}
\label{fig3}
\end{figure}

Note that the components $E_r$ leads both $E_{\theta}$ and $E_z$ by $\pi/2$. The same phase difference exists between $B_{\theta}$ and $B_r$, with $B_r$ leading. Equations (\ref{Er})-(\ref{Btheta}) may be written more compactly, using the expressions found above for $\omega$, $k_z$, and $v_g$. 

\begin{figure}[t]
\centering
\includegraphics[width=8cm]{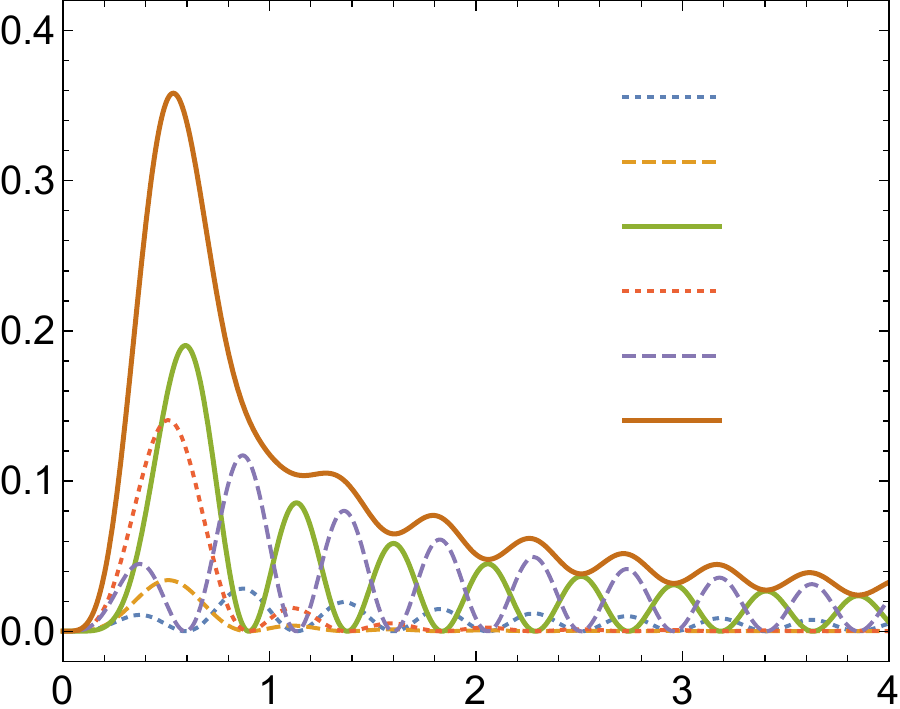}
\begin{picture}(0,0)(0,0)
\put(-118,-13){$r/\lambda_0$}
\put(-48,153){$|E_r/E_0|^2$}
\put(-48,136){$|E_\theta/E_0|^2$}
\put(-48,120){$|E_z/E_0|^2$}
\put(-48,103){$|cB_r/E_0|^2$}
\put(-48,87){$|cB_\theta/E_0|^2$}
\put(-40,71){$Sum$}
\put(-130,130){$(b) ~l = 3$}
\end{picture}
\vskip4mm
\caption{Same as Fig.\ref{fig3}, but for the case of $l = 3$, using parameters the same as in Fig. \ref{fig1}.}
\label{fig4}
\end{figure}
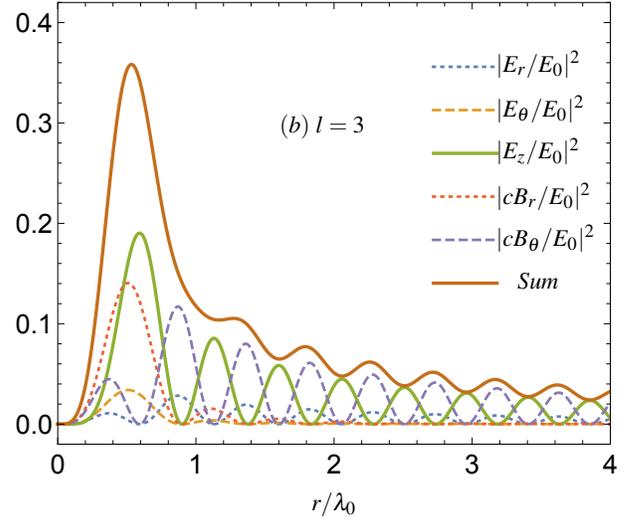

Density plots of the electric and magnetic field intensity profiles are shown in Fig. \ref{fig1}, for $l = 1$ and $l = 3$. The fact that the density plots in the {\it moving focal plane}, at $t = 1$ fs, are identical to their counterparts in the {\it focal plane} at $t = 0$ in Ref. \cite{salamin-sr} is consistent with the earlier conclusion that the bullet propagates undistorted in the under-dense plasma. Note that profile of a negative-order field component is the same as that of the corresponding positive-order component, due to the relation $J_{-l} = (-1)^l J_l$. This, however, is not the case for the spiral phases shown in Fig. \ref{fig2}, for $l = \pm1$ and $l = \pm3$. Phases corresponding to $+l$ and $-l$ have opposite handedness. 

Intensity profiles of all the components given by Eqs. (\ref{Er})-(\ref{Btheta}) together with their sums are shown in Figs. \ref{fig3} and \ref{fig4}, for the cases of $l = 1$ and $l = 3$, respectively, as functions of the radial distance from the moving focus. Relative brightness of the rings displayed in Fig. \ref{fig1} may be better seen by comparing the heights of the corresponding peaks in Figs. \ref{fig3} and \ref{fig4}. Note that ``Sum" stands for the sum of all the other intensity profiles, while it also represents the scaled energy density $u/u_0$, where $u$ is given by Eq. (\ref{u}) below, and $u_0 = \varepsilon_0E_0^2/2$.

\section{Time-averaged densities}

In this section, expressions will be derived for the time-averaged densities of several physical quantities pertaining to a Bessel-Bessel bullet, employing the fields (\ref{Er})-(\ref{Btheta}). Such expressions will ultimately be needed in applications for which the fields may be of utility \cite{mcdonald1,allen1,allen2}.

The field components (\ref{Er})-(\ref{Btheta}) have quasi-harmonic time-dependence. Since, in the moving focal plane, $\zeta = 0$ and $\eta = ct$, dependence upon the time is of the form $e^{-i\omega't}$, or at an effective frequency $\omega' = \alpha c$. Thus, the time-average $\langle X\rangle$ of a quantity $X(t)$, expressible as the product of two quantities, $Y(t)$ and $Z(t)$, will be found from $\langle X \rangle = (YZ^*+Y^*Z)/2$ \cite{jackson}.

\subsection{Energy}

After some algebra, the time-averaged electromagnetic energy density in the ultrashort and tightly-focused pulse may be cast in the form
\begin{eqnarray}
\label{u}	\langle u \rangle &=& \frac{1}{2}\varepsilon_0(|E|^2+|cB|^2),\nonumber\\
	   &=&\frac{u_0}{(\alpha+2k_0)^2}\left\{(\alpha^2+4k_0^2)\left(\frac{k_r}{k_0}\right)^2 \left[J_{l+1}^2+J_{l-1}^2\right]\right.\nonumber\\
	   & & \left. +16\alpha^2 \left[1-\frac{(\pi/L)^2}{3k_0(\alpha+2k_0)}\right]^2J_l^2\right\} .
\end{eqnarray}

\begin{figure}[t]
\centering
\includegraphics[width=8cm]{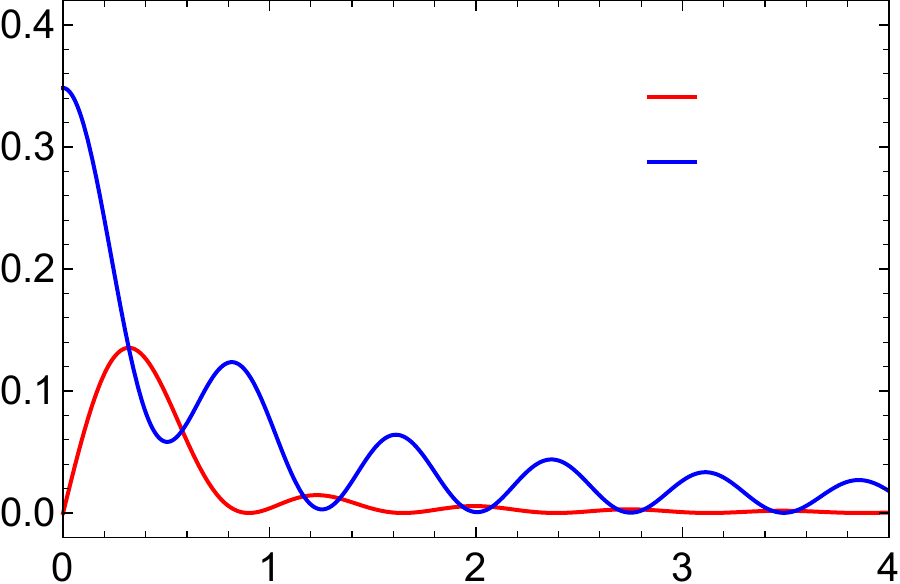}
\begin{picture}(0,0)(0,0)
\put(-117,-13){$r/\lambda_0$}
\put(-49,121){$\langle cp_\theta/u_0\rangle$}
\put(-49,105){$\langle cp_z/u_0\rangle$}
\put(-130,120){$(a) ~l = 1$}
\end{picture}
\vskip4mm
\caption{Time-averaged linear momentum components at points in the moving focal plane, as a function of the radial distance from the focus, for the case of $l = 1$, and employing the same parameters as in Fig. \ref{fig1}.}
\label{fig5}
\end{figure}
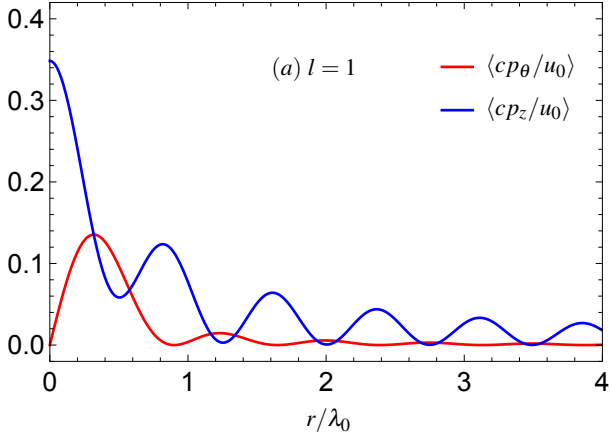

In the elaborate algebra leading to Eq. (\ref{u}) the transformation $(2l/\rho) J_l(\rho) = J_{l+1}(\rho)+J_{l-1}(\rho)$ has been used repeatedly. General agreement between this expression and its vacuum, long-pulse, counterpart  \cite{mcdonald1} may be seen by letting $k_p\to0$ and $L\to\infty$. 

\subsection{Linear momentum}

The electromagnetic linear momentum density is given by $\bm{p} = \varepsilon_0(\bm{E}\times \bm{B})$. In cylindrical coordinates, with unit vectors $\hat{\bm{r}}$, $\hat{\bm{\theta}}$, and $\hat{\bm{z}}$, its time-averaged value is 
\begin{eqnarray}\label{p}
	\langle \bm{p} \rangle &=& \frac{\varepsilon_0}{2} (\bm{E}\times \bm{B}^*+\bm{E}^*\times \bm{B}),\nonumber\\
	                       &=& \frac{u_0}{c} \left[\frac{\alpha-2k_0}{\alpha+2k_0}\right] \left\{\left(\frac{8l\alpha/k_0}{\alpha-2k_0}\right)\right.\nonumber\\
	                       & &\left.\times
	              \left[1-\frac{(\pi/L)^2}{3k_0(\alpha+2k_0)}\right] \left[\frac{J_l^2}{r}\right] \hat{\bm{\theta}}\right.\nonumber\\
	              & &\left. 
	              -\left(\frac{k_r}{k_0}\right)^2\left[J_{l+1}^2+J_{l-1}^2\right]\hat{\bm{z}}\right\}.
\end{eqnarray}

Due to the fact that the fields in (\ref{Er})-(\ref{Btheta}) are not purely transverse, the pulse carries forward, as well as azimuthal, linear momentum, according to Eq. (\ref{p}). Note, however, that to the integration of $\langle \bm{p} \rangle$ over a plane at a fixed $z$, only the $z-$component contributes. Combined with the absence of a radial component from the linear momentum density expression, this demonstrates that the pulse does not spread transversely.

\begin{figure}[t]
\centering
\includegraphics[width=8cm]{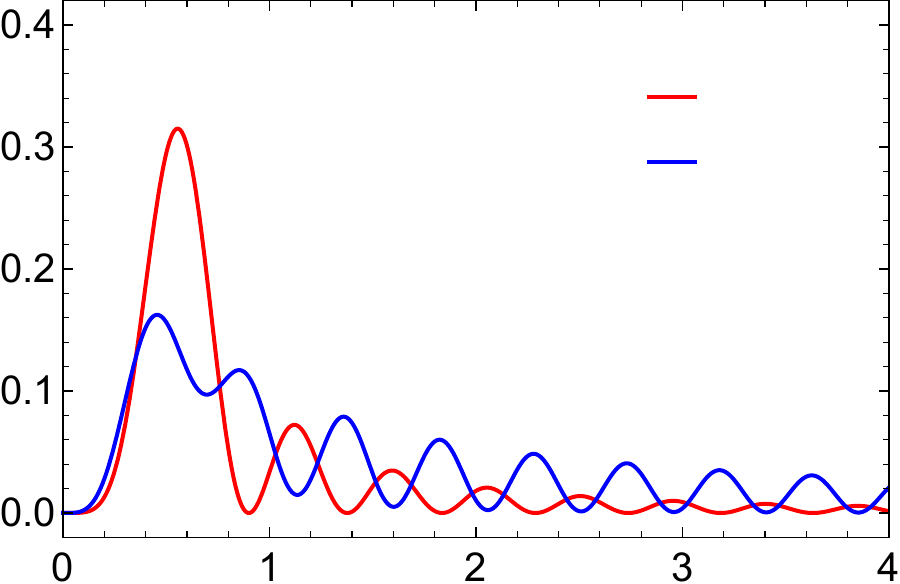}
\begin{picture}(0,0)(0,0)
\put(-117,-13){$r/\lambda_0$}
\put(-49,121){$\langle cp_\theta/u_0\rangle$}
\put(-49,105){$\langle cp_z/u_0\rangle$}
\put(-130,120){$(b) ~l = 3$}
\end{picture}
\vskip4mm
\caption{Same as Fig. \ref{fig5}, but for the case of $l = 3$, and employing the same parameters as in Fig. \ref{fig1}.}
\label{fig6}
\end{figure}
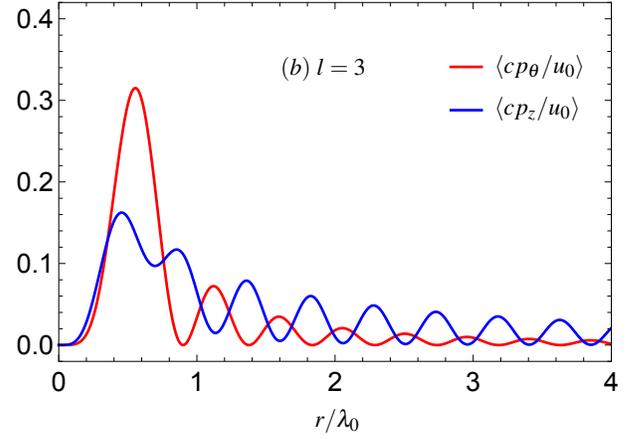

Variations of $\langle p_\theta\rangle$ and $\langle p_z\rangle$ in the moving focal plane, with the radial distance from the moving focus, for the cases of $l = 1$ and $l = 3$, are shown in Figs. \ref{fig5} and \ref{fig6}, respectively. A peak in these plots marks the radius of the center of a ring of maximum linear momentum, in the corresponding density plot. Note that all density plots would be hollow, apart from that of $\langle p_z\rangle$ of the $l = 1$ case.

\subsection{Radiation intensity and power}

The Poynting vector, representing the energy flux density, is $\bm{S} = \varepsilon_0 c^2\bm{E}\times\bm{B} = c^2 \bm{p}$. Hence, the time-averaged electromagnetic energy flux density of the pulse is $\langle \bm{S} \rangle = c^2 \langle \bm{p} \rangle$, which follows from Eq. (\ref{p}). The axial component, $\langle S_z \rangle $, gives the intensity of the pulse, in W/m$^2$, as a function of the radial distance $r$
\begin{equation}\label{intensity}
	I(r) = cu_0 \left[\frac{k_0-\alpha/2}{k_0+\alpha/2}\right] \left(\frac{k_r}{k_0}\right)^2 [J_{l+1}^2+J_{l-1}^2].
\end{equation}

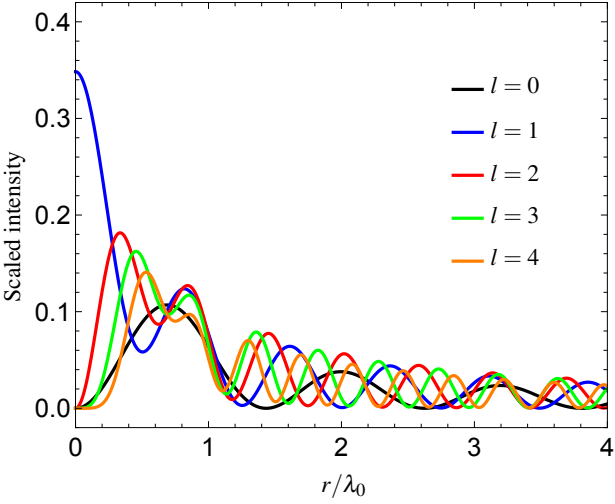
\begin{figure}[t]
\centering
\includegraphics[width=7.7cm]{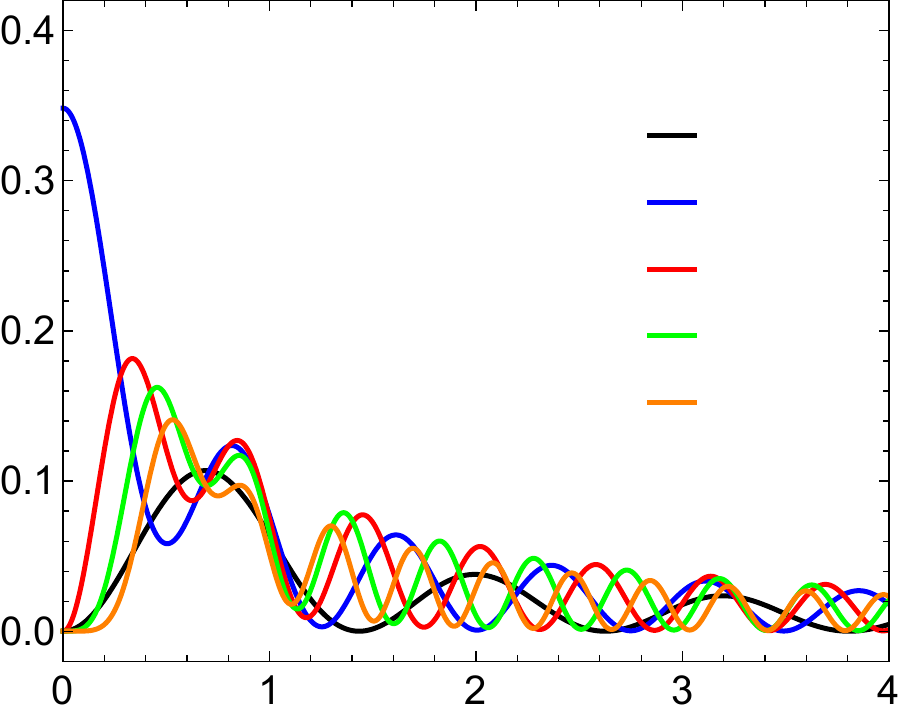}
\begin{picture}(0,0)(0,0)
\put(-113,-13){$r/\lambda_0$}
\put(-233,59){\begin{sideways}Scaled intensity\end{sideways}}
\put(-49,138){$l = 0$}
\put(-49,121){$l = 1$}
\put(-49,105){$l = 2$}
\put(-49,89){$l = 3$}
\put(-49,73){$l = 4$}
\end{picture}
\vskip4mm
\caption{Scaled intensity $I/cu_0$ in the focal plane as a function of the distance from focus, for the parameters of Fig. \ref{fig1}.}
\label{fig7}
\end{figure}

A density plot of $I$ in the moving focal plane would exhibit alternating bright and dark rings.  In lieu of density plots, however, variations of the scaled intensities in the moving focal plane are shown in Fig. \ref{fig7}, as functions of the radial distance from focus, for the cases corresponding to $l = 0, 1, \cdots, 4$.  Here, too, the rings are all hollow, except for the $l = 1$ case. 

Finally, one gets an expression for the power carried by the pulse by integrating $I(r)$ over the moving focal plane
\begin{equation}\label{power}
	P = 2\pi cu_0 \left[\frac{k_0-\alpha/2}{k_0+\alpha/2}\right] \left(\frac{k_r}{k_0}\right)^2\int_0^{\bar{r}}[J_{l+1}^2+J_{l-1}^2]rdr,
\end{equation}
in which $\bar{r} \gtrsim w_0$ is a measure of the transverse spatial extension of the pulse.

Note at this point that, like the wavevector (\ref{k}) lines of flow of the linear momentum and energy flux vectors are not parallel to the direction of propagation, but follow helices of fixed radii \cite{mcdonald1,berry}. With $\langle \bm{p}\rangle = \langle p_\theta\rangle\hat{\bm{\theta}}+\langle p_z\rangle \hat{\bm{z}}$, both vectors make the angle $\gamma$ with the direction of propagation, given by  $\tan\gamma = \langle p_\theta\rangle/\langle p_z\rangle$. 

\subsection{Angular momentum}

In cylindrical coordinates, the position vector of a point within the pulse is $\bm{r} = r\hat{\bm{r}}+z\hat{\bm{z}}$. Thus, the angular momentum density is $\bm{l}= \bm{r}\times\bm{p} =\varepsilon_0 \bm{r}\times(\bm{E}\times\bm{B})$. Hence, the time-averaged angular momentum density may be cast in the following form
\begin{eqnarray}
	\label{l} \langle \bm{l} \rangle &=& \frac{\varepsilon_0}{2} \bm{r}\times(\bm{E}\times\bm{B}^*+\bm{E}^*\times\bm{B}),\nonumber\\
	 &=& u_0 \left[\frac{\alpha-2k_0}{\alpha+2k_0}\right] \left\{\left(\frac{8\alpha}{\alpha-2k_0}\right)  \left[1-\frac{(\pi/L)^2}{3k_0(\alpha+2k_0)}\right] 
	              \right.\nonumber\\
	  & &\left.\times J_l^2 \left[-\left(\frac{l}{\omega_0}\right)\left(\frac{z}{r}\right)\hat{\bm{r}}+\left(\frac{l}{\omega_0}\right)\hat{\bm{z}}\right]\right.\nonumber\\
	  & &\left. +\left(\frac{k_r}{k_0}\right)^2\left[J_{l+1}^2+J_{l-1}^2\right] \left(\frac{r}{c}\right) \hat{\bm{\theta}}\right\},
\end{eqnarray}
where $\omega_0 = ck_0$. Writing $\langle \bm{l} \rangle = \langle l_r \rangle\hat{\bm{r}}+\langle l_\theta \rangle\hat{\bm{\theta}}+\langle l_z \rangle\hat{\bm{z}}$, contribution to the integration of the angular momentum density over a plane of fixed $z$ (perpendicular to the direction of propagation) comes only from $\langle l_z \rangle$. Thus, $\langle l_z \rangle $ is the time-averaged density of orbital angular momentum about the direction of propagation \cite{mcdonald1,berry}. Note that $\langle l_z \rangle = 0 $, for $l=0$, as expected. It is also worth pointing out the striking resemblance of Eq. (\ref{l}) for the angular momentum of a Bessel-Bessel bullet, to Eq. (8) in Ref. \cite{allen1} for the angular momentum of the Laguerre-Gaussian laser modes \cite{allen2}.

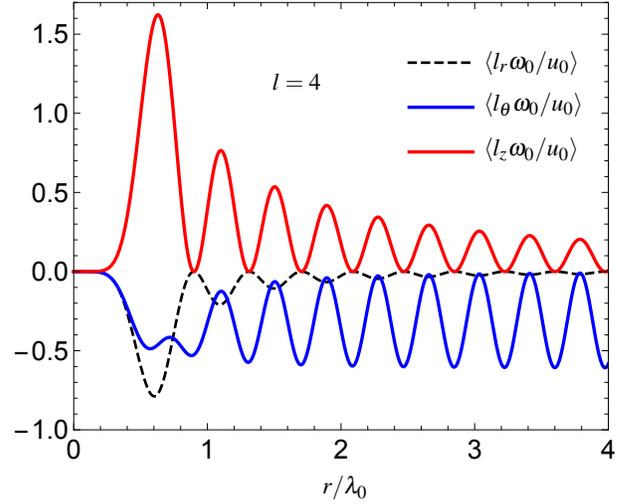
\begin{figure}[t]
	\centering
	\includegraphics[width=8cm]{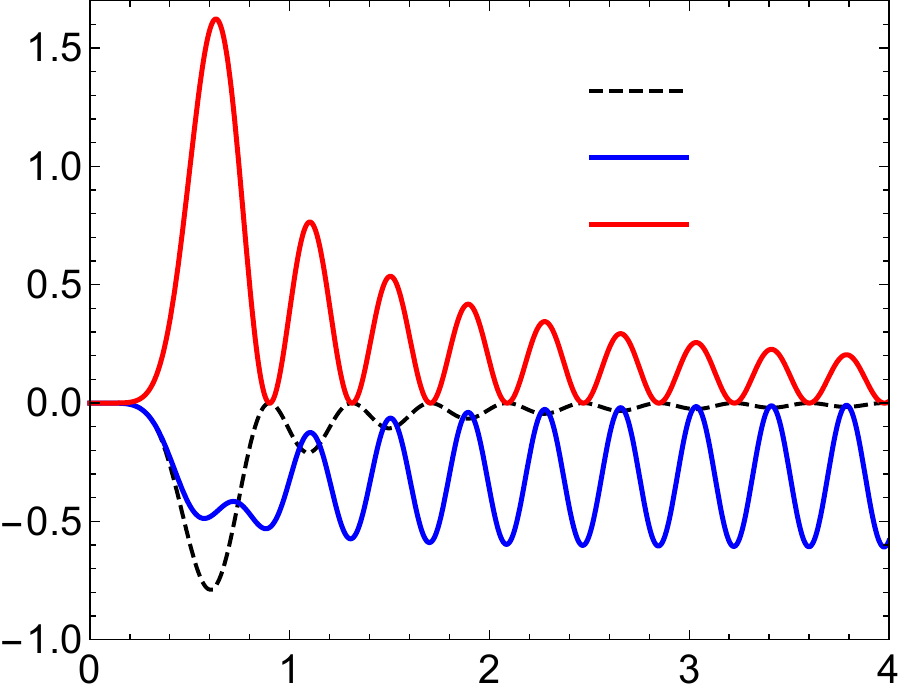}
	\begin{picture}(0,0)(0,0)
	\put(-113,-13){$r/\lambda_0$}
	\put(-52,148){$\langle l_r\omega_0/u_0\rangle$}
	\put(-52,132){$\langle l_\theta\omega_0/u_0\rangle$}
	\put(-52,115){$\langle l_z\omega_0/u_0\rangle$}
	\put(-132,140){$l = 4$}
	\end{picture}
	\vskip4mm
	\caption{Components of time-averaged angular momentum density, scaled by $u_0/\omega_0$, in the focal plane as a function of the distance from focus, for parameters the same as in Fig. \ref{fig1}.}
	\label{fig8}
\end{figure}

As an example, variation with the radial distance from the moving focus, of the time-averaged components of the angular momentum density, are shown in Fig. \ref{fig8}, for the case of $l = 4$. It can be inferred from this figure that the corresponding density plots would all consist of hollow concentric rings. 

To understand the behaviour of the different components away from the focal point, one may use the asymptotic representation 
\begin{equation}
	J_l(k_rr) \sim \sqrt{\frac{2}{\pi k_rr}} \cos\left(k_rr-\frac{l\pi}{2}-\frac{\pi}{4}\right).
\end{equation}
The oscillations are obviously due to the $\cos$ function. It is easy also to see that, asymptotically, $l_r \sim-r^{-2}$, which explains why $\langle l_r\omega_0/u_0\rangle$ decays quickly to zero away from the focus, while $l_z \sim r^{-1}$ causes $\langle l_z\omega_0/u_0\rangle$ to decay to zero slowly by comparison. Finally, what appears to be an oscillation between the same maximum and minimum value of $\langle l_\theta\omega_0/u_0\rangle$ is due to the fact that $l_\theta$ is asymptotically independent of $r$. Asymptotic behaviour of the quantities shown in Figs. \ref{fig3}-\ref{fig7} may be understood on the basis of considerations similar to the ones just outlined for Fig. \ref{fig8}.

\section{Summary and conclusions}

Fields of an ultra-short and tightly-focused laser Bessel pulse have recently been derived analytically, for the first time \cite{salamin-oe,salamin-sr}. The pulse, dubbed a Bessel-Bessel bullet, has been shown to propagate without dispersion or diffraction inside an under-dense plasma. For the derived fields to be of utility in potential applications, they have here been supported by further investigation of some of their key propagation characteristics, along with important time-averaged quantities pertaining to them. It has been shown that a Bessel-Bessel bullet, propagating in an under-dense plasma, carries electromagnetic linear and angular momenta. Analytic expressions have been derived for the time-averaged energy density, linear momentum density, energy flux density, and angular momentum density. It has further been shown that the bullet possesses orbital angular momentum about its direction of propagation.

\section*{acknowledgements}

The author thanks K. Z. Hatsagortsyan for fruitful discussions and a critical reading of the manuscript.

\end{document}